\documentclass[11pt]{article}

\usepackage[preprint]{acl}

\usepackage{times}
\usepackage{latexsym}

\usepackage[T1]{fontenc}

\usepackage[utf8]{inputenc}

\usepackage{microtype}
\usepackage{inconsolata}

\usepackage{amsmath}

\usepackage{graphicx}
\usepackage{algorithm}
\usepackage{algorithmic}

\title{\textsc{MARLaaS}: Multi-Tenant Asynchronous Reinforcement \\ Learning as a Service}

\author{
  \textbf{Timothy Tin Long Yu}\textsuperscript{1}\thanks{Equal contribution.},
  \textbf{Gursimran Singh}\textsuperscript{1}\footnotemark[1],
  \textbf{Ge Shi}\textsuperscript{1},
  \textbf{Hanieh Sadri}\textsuperscript{1},
  \\
  \textbf{Yong Zhang}\textsuperscript{1},
  \textbf{Zhenan Fan}\textsuperscript{1}\thanks{Correspondence: zhenan.fan1@huawei.com}
  \\
  \textsuperscript{1}Huawei Technologies Canada 
  \\
}

\usepackage{eso-pic}
\usepackage{graphicx}

\AddToShipoutPictureBG*{%
  \AtPageUpperLeft{%
    \raisebox{-2cm}{%
      \hspace{0.9\paperwidth}%
      \includegraphics[width=0.085\textwidth]{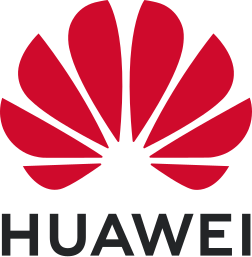}%
    }%
  }%
}

\begin{document}
\maketitle
\begin{abstract}
Reinforcement Learning from Verifiable Rewards (RLVR) has significantly improved the reasoning capabilities of large language models (LLMs), particularly in multi-turn agentic settings involving environment interaction like tool use. However, fine-tuning such models remains prohibitively expensive due to high computational requirements, limiting accessibility.
We propose \textbf{\textsc{MARLaaS}} (Multi-tenant Asynchronous RL as a Service), a system for concurrent RL fine-tuning across multiple users and tasks. Our approach is based on two key ideas: (1) sharing a base model across tenants using lightweight LoRA adapters, and (2) a disaggregated asynchronous architecture that decouples rollout generation, environment interaction, and policy training into independently scheduled stages. This design enables tasks to progress through the RL pipeline at their own pace in an event-driven manner, reducing cross-task interference, idle time, and end-to-end latency.
In multi-task settings (we report up to 32 concurrent tasks), MARLaaS achieves single-task state-of-the-art performance while improving accelerator utilization by up to 4.3× and reducing end-to-end training time by 85\%.
\end{abstract}

\section{Introduction}

\begin{figure}[t]
    \centering
    \includegraphics[width=\columnwidth]{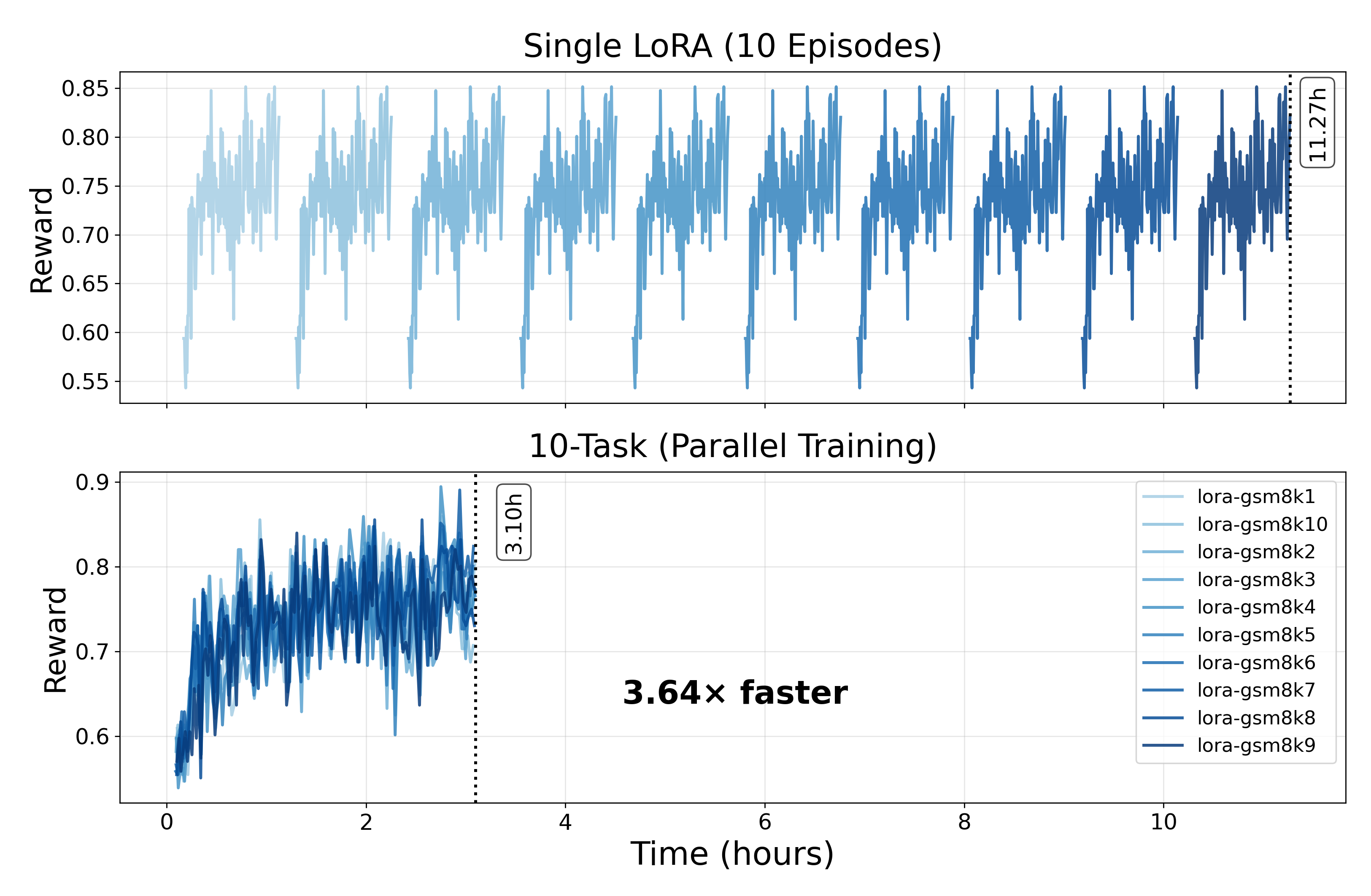}
    \caption{\textbf{Scaling RLVR to 10 concurrent LoRA tasks.} \textsc{MARLaaS} maintains stable reward improvements under high multi-tenant load, demonstrating superior scalability to single-disaggregated baselines over one epoch.}
    \label{fig:teaser}
\end{figure}

Large language models (LLMs) have demonstrated strong reasoning capabilities, particularly when fine-tuned using reinforcement learning with verifiable rewards (RLVR), which trains models using objective reward signals from verifiable outcomes such as correctness of solutions, passing unit tests, or successful tool execution \cite{kaufmann2024surveyreinforcementlearninghuman}. 

RLVR training alternates between three tightly coupled phases: \textbf{(1) rollout generation}, where the current policy produces candidate responses; \textbf{(2) environment interaction}, including tool calls, external API invocations, and reward computation; and \textbf{(3) policy training}, where gradient updates are applied based on collected trajectories. In agentic settings, these phases may be invoked multiple times per episode as the model learns to decompose complex problems into manageable sub-tasks \cite{wang2023voyageropenendedembodiedagent}.

In practice, the three distinct phases happen sequentially and block other phases from proceeding. Specifically, during training, rollout workers sit idle waiting for the new policy; when tool calls are dispatched, accelerators wait for external responses; when rollouts generate tokens, training resources remain unused. These idle periods waste valuable compute and drive up costs. 
A second dimension arises from per-tenant resource allocation where each task must instantiate its model and take up substantial memory. As task concurrency increases, this overhead becomes prohibitive, limiting scalability and increasing per-user costs.

Consequently, RLVR training pipelines are computationally and memory expensive, often requiring large clusters of accelerators for extended training runs. This makes RLVR inaccessible to many researchers and practitioners, motivating the need for a cost-efficient RL Training-as-a-Service (RLTaaS) platform.

One line of work in the literature tackles these inefficiencies by overlapping different phases of the RL pipeline. Existing RL fine-tuning systems address pipeline inefficiencies through two main approaches. AReaL \cite{fu2025areallargescaleasynchronousreinforcement} introduces \emph{staleness}—allowing \textit{off-policy updates} using rollouts from previous policy versions—enabling overlap between rollout and training phases. While this increases throughput, recent findings \cite{li2026unleashingefficientasynchronousrl} show that excessive staleness degrades convergence quality. 
SkyRL Agent \cite{cao2025skyrlagentefficientrltraining} enables rollout generation to overlap with tool calls, reducing accelerator idling during environment interactions. However, both approaches are designed for single-task training and naive extensions to multi-tenant settings either replicate resources per task, reintroducing prohibitive overhead, or execute tasks sequentially, leading to underutilization. 

End users prioritize stable training and high-quality final models over marginal reductions in training time. In contrast, cloud providers are incentivized to maximize hardware utilization, as resource inefficiencies directly translate into higher operational costs that must be reflected in user pricing. An competitive RLTaaS design must therefore improve resource utilization while preserving training quality, avoiding mechanisms that introduce harmful distribution shift or destabilize policy optimization.

We present \textbf{\textsc{MARLaaS}}, a multi-task asynchronous RL fine-tuning system designed to maximize hardware accelerator utilization by co-training multiple tasks on the same hardware and overlapping their RLVR phases—training, rollout, and environment interactions—across various tasks. Specifically, our approach builds on two key principles: \textbf{(1)} sharing a base model across tenants via lightweight LoRA adapters to minimize per-task memory overhead and support cross-task batching of rollouts; and \textbf{(2)} a disaggregated asynchronous architecture that decouples rollout generation, environment interaction, and policy training, allowing each task to progress with minimal cross-task interference. Unlike prior asynchronous RL systems, \textsc{MARLaaS} preserves strict per-task policy consistency by ensuring that each task trains only on trajectories generated from its latest committed policy version, without relying on stale rollouts or off-policy updates. Rather than introducing intra-task policy staleness, \textsc{MARLaaS} achieves asynchrony through cross-task phase overlap, where rollout, environment interaction, and training from different tasks execute concurrently without global synchronization.

%

Further, our design is motivated by a key insight: RLVR workflow is typically rollout-bound, with rollout generation and environment interactions dominating computation time (10-100× longer than training updates), especially in agentic scenarios with tool calls. We design \textsc{MARLaaS} to counter this asymmetry by allocating resources such that rollout throughput can be maximized. 
Specifically, it uses multi-LoRA batching during rollout generation to maximize throughput at the bottleneck through multi-task concurrency, while policy training across tasks are intentionally performed sequentially to minimize per-task training latency and enable rapid phase transitions. This asymmetric design keeps hardware continuously utilized through natural task interleaving rather than staleness.

\subsection{Contributions}

Our key contributions are as follows:

\textbf{(1)} We study multi-tenant RLVR training and introduce a system for RL Training-as-a-Service that enables efficient training of multiple tasks using shared resources.

\textbf{(2)} We identify a rollout-training asymmetry in RLVR workloads, where rollout generation and environment interaction dominate execution time, creating a rollout-bound bottleneck. We show that this bottleneck can be alleviated through cross-task rollout concurrency, which is orthogonal to staleness-based approaches and provides an additional dimension for improving utilization.

\textbf{(3)} We design a phase-aware asynchronous scheduling system that decouples rollout generation, environment interaction, and training, enabling independent progression of concurrent tasks with minimal cross-task interference.

\textbf{(4)} We empirically demonstrate that \textsc{MARLaaS} improves accelerator utilization by up to 4.3\textbf{x} and increasing total training throughput by up to 5.7\textbf{x} while maintaining reward performance across multi-task settings, achieving comparable rewards to single-task baselines in up to 85\% less time.






\section{Related Work}

\subsection{Reinforcement Fine-Tuning}
RLVR has become a key approach to improving LLM reasoning and alignment \cite{ouyang2022traininglanguagemodelsfollow, schulman2017proximalpolicyoptimizationalgorithms, rafailov2024directpreferenceoptimizationlanguage, openai2024gpt4technicalreport, guo2025deepseek, bai2022traininghelpfulharmlessassistant}. Prior works focus on improving optimization stability and sample efficiency through algorithmic advances in policy optimization and reward modeling. These approaches are complementary to our work, which addresses system-level challenges in scaling RLVR to multi-tenant settings.

\subsection{RLVR Training Systems}
RLVR training systems differ in system designs. \textbf{Synchronous systems} (e.g., VERL \cite{Sheng_2025}) enforce strictly sequential phase separation leading to resource bubbles and underutilization. \textbf{Asynchronous systems} (e.g., \cite{fu2025areallargescaleasynchronousreinforcement}) overlap the phases by introducing policy staleness and off-policy bias that may negatively impact final model convergence \cite{li2026unleashingefficientasynchronousrl}. Other systems overlap rollout generation with environment interactions (e.g., \cite{cao2025skyrlagentefficientrltraining}) or optimizing pipeline scheduling, workload balancing, and rollout latency \cite{sheng2025laminarscalableasynchronousrl, noukhovitch2025asynchronousrlhffasterefficient, wu2025llamarldistributedasynchronousreinforcement, gao2025rollpackermitigatinglongtailrollouts}.

While these systems target the efficiency of RLVR, they primarily target \emph{single-task} training. Extending them to multi-tenant settings either leads to sequential execution and poor utilization or requires duplicating infrastructure per task, incurring high memory and compute costs. Conversely, we leverage LoRAs to efficiently perform multi-task shared execution across concurrent RL workloads.

On the other hand, Tinker~\cite{skyrl2026tinker} and SkyRL-Train define a minimal post-training API that decouples algorithmic logic from infrastructure, enabling portability for multi-task RLVR execution and supporting multi-LoRA workflows. However, these efforts remain largely at the API level and do not provide concrete system designs or implementations. In particular, they do not address scheduling, resource sharing, or pipeline orchestration required to efficiently support multi-tenant RLVR workloads.

\subsection{Multi-LoRA Systems}
LoRA enables parameter-efficient fine-tuning via low-rank updates to shared model weights \cite{hu2021loralowrankadaptationlarge}. Prior work has explored multi-tenant LoRA serving for both inference \cite{sheng2024sloraservingthousandsconcurrent, chen2023punicamultitenantloraserving} and training \cite{ye2024mlorafinetuningloraadapters, li2026tloraefficientmultiloratraining, zuo2026altoadaptiveloratuning, Lin_2025}, primarily focusing on improving throughput by batching multiple adapters.

In contrast, \textsc{MARLaaS} adopts an asymmetric design tailored to RLVR workloads. We leverage multi-LoRA batching during rollout generation to improve throughput in the dominant phase, while keeping training updates lightweight to maintain low latency and enable efficient phase interleaving across tasks.

\subsection{Summary of Gaps}
Prior work addresses either (i) single-task RL scaling through asynchronous execution and staleness, (ii) multi-LoRA efficiency for inference or supervised training, or (iii) agent training frameworks for individual tasks. None jointly consider multi-tenant RLVR training, where rollout generation dominates compute (10-100× more than training), and multiple on-policy tasks sharing resources. \textsc{MARLaaS} fills this gap by exploiting rollout-training asymmetry through asymmetric multi-LoRA application and phase-aware multi-task scheduling.

\begin{figure}[t]
    \centering


    \includegraphics[width=0.8\linewidth]{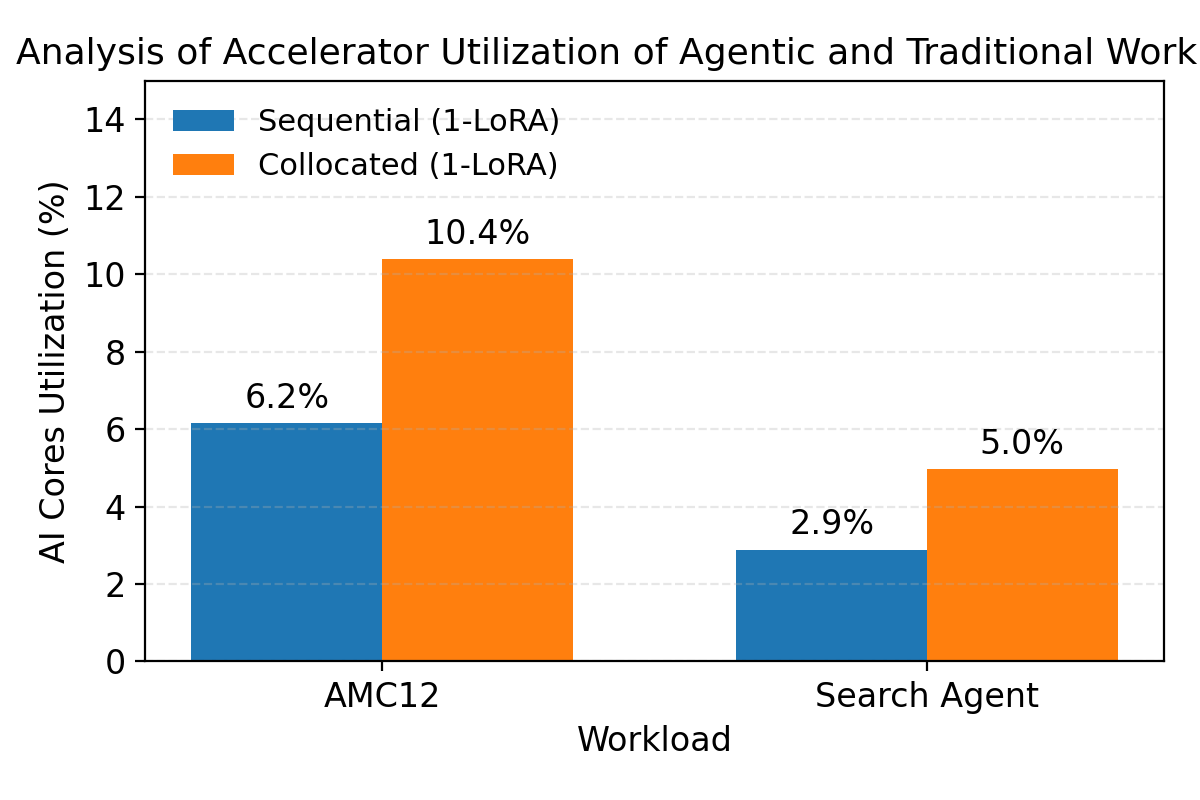}
    \caption{
    Accelerator utilization under naive multi-tenant RL training-as-a-service strategies across traditional (AMC12) and agentic search workloads. Both Single-Disaggregated and Single-Collocated baselines exhibit significant under-utilization, particularly in agentic settings where tool and environment latency introduce irregular rollout delays.
    }
    \label{fig:under-util}
\end{figure}

\begin{figure*}[t]
    \centering
    \includegraphics[width=0.9\textwidth]{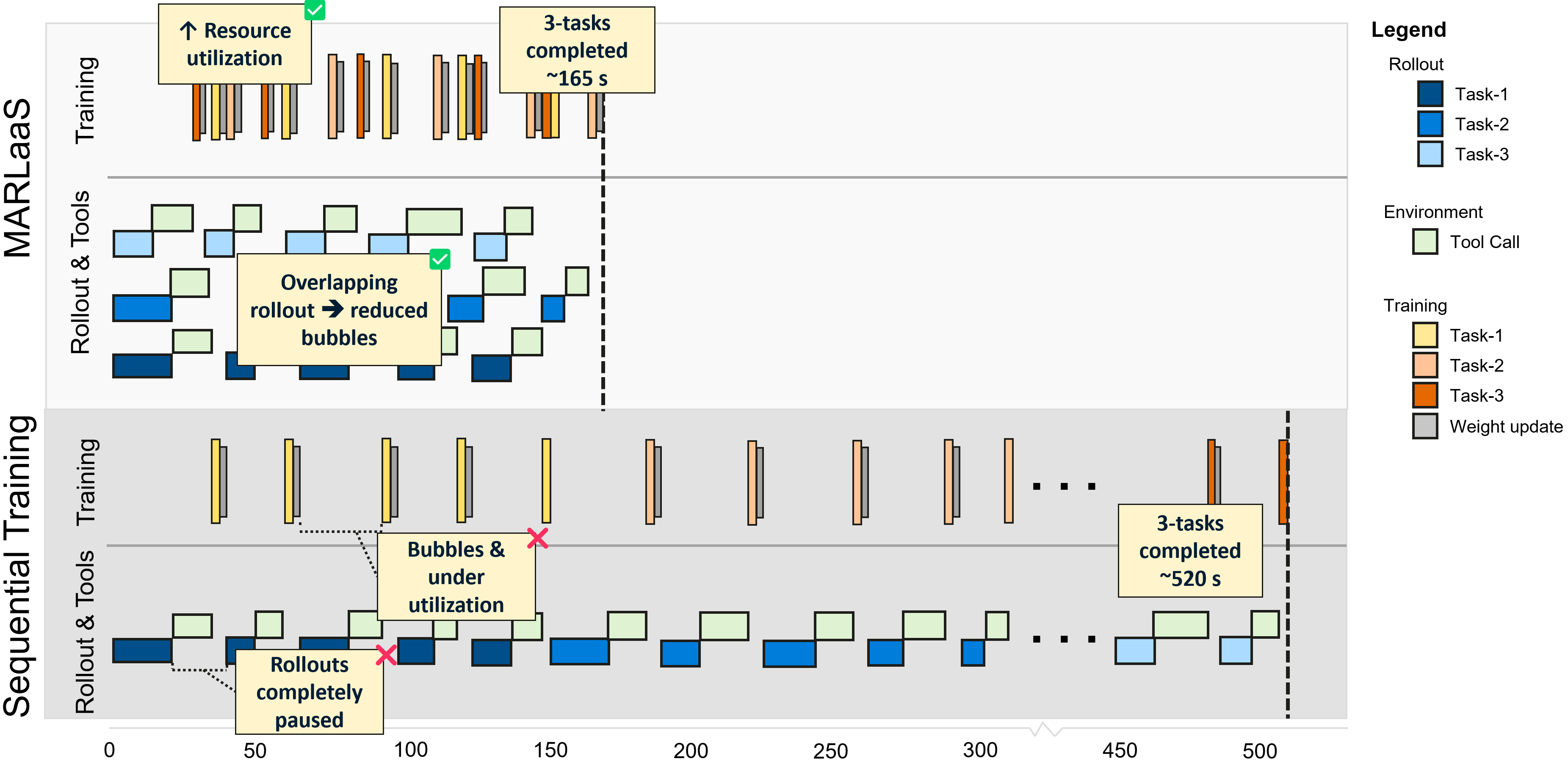}
    \caption{
    \textbf{Execution timeline of \textsc{MARLaaS} compared to a Single-Disaggregated baseline for three-task RL fine tuning.} 
    We present a graphic that demonstrates training 3 Qwen3-0.6B models on an agentic search workload. Training (warm colors), rollout (cool colors), and environment tool calling (green) phases are shown assuming all tasks are submitted at $t=0$. By enabling asynchronous phase transitions and batching rollouts across heterogeneous tasks, \textsc{MARLaaS} reduces pipeline bubbles and idle GPU time.
    }
    \label{fig:execution-timeline}
\end{figure*}

\section{Motivation: Inefficiency and Asymmetry in RLVR Workloads}

\begin{figure}[t]
    \centering
    \includegraphics[width=\linewidth]{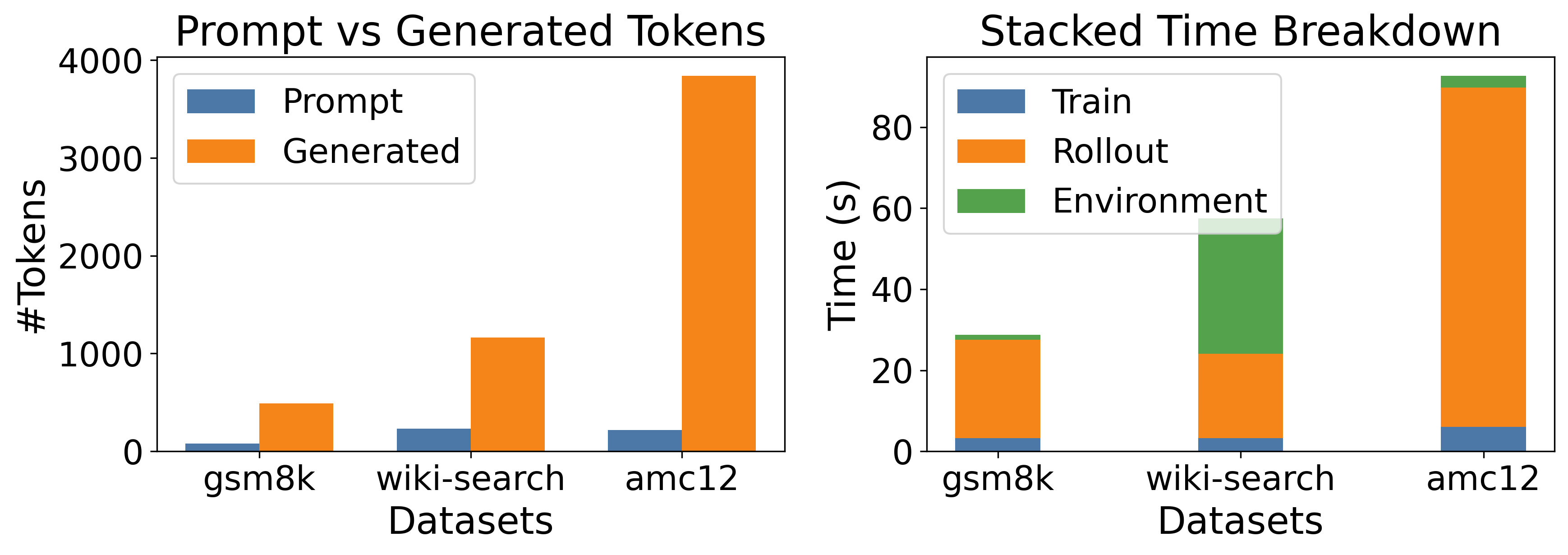}
    \caption{Token usage and runtime breakdown across tasks. Rollout and environment interaction dominate execution time, especially in agentic workloads.}
    \label{fig:nooftokens-and-timebreakdownb}
\end{figure}
\textbf{RLVR workloads are dominated by rollout and environment interaction.}
As shown in Figure~\ref{fig:nooftokens-and-timebreakdownb}, RLVR workloads exhibit a strong imbalance between rollout generation, environment interaction, and policy training. Across both reasoning and agentic tasks, rollout generates substantially more decode tokens than prompt input tokens for prefill (Figure~\ref{fig:nooftokens-and-timebreakdownb}, left), and dominates total execution time (Figure~\ref{fig:nooftokens-and-timebreakdownb}, right), often exceeding policy update latency by an order of magnitude.

This imbalance is especially pronounced in agentic settings, where tool use introduces additional latency and variability.

\textbf{Sequential execution creates pipeline inefficiency.}
As illustrated in Figure~\ref{fig:execution-timeline}, the sequential coupling between phases results in significant pipeline bubbles. Existing RLVR systems execute rollout, environment interaction, and training sequentially within each task. Idle time occurs when rollout workers wait for training updates, training workers wait for rollout completion, and both remain idle during environment interaction. As a result, accelerators are significantly underutilized by phase interference rather than raw compute limits. 

\textbf{Inefficiencies of multi-tenant heterogeneous pacing under shared execution.}
In multi-tenant settings, different tasks progress through the RL pipeline at significantly different rates due to variation in rollout complexity and environment interaction latency. When sharing resources across tasks, this heterogeneity leads to misaligned compute demand across tasks, resulting in resource utilization bubbles.

Table~\ref{tab:rollout-tradeoff} illustrates the synchronization overhead introduced when jointly training heterogeneous RLVR workloads. We concurrently train three tasks with distinct workload characteristics (\textsc{GSM8K}, wiki-search, and \textsc{AMC12}) and measure rollout latency together with the induced waiting time under synchronized multi-task training.

As shown, rollout latencies vary substantially across tasks due to differences in reasoning complexity and environment interaction. Under synchronized execution, faster tasks must idle while waiting for slower rollouts to complete, leading to considerable synchronization overhead and reduced hardware utilization. This effect becomes particularly pronounced for heterogeneous workloads, where slow-moving tasks can dominate overall iteration time despite representing only a subset of the active workloads.

These results motivate asynchronous execution designs that decouple task progression from global synchronization barriers, enabling tasks to independently progress through rollout and training phases as resources become available.

\begin{table}[t]
\centering
\caption{
Rollout latency and synchronization-induced waiting time under synchronized multi-task RLVR training. We jointly train three heterogeneous tasks (\textsc{GSM8K}, wiki-search, and \textsc{AMC12}) and measure the rollout latency of each task together with the waiting time introduced by global synchronization.
}
\small
\label{tab:rollout-tradeoff}
\setlength{\tabcolsep}{5pt}
\begin{tabular}{lccc}
\hline
\textbf{Task} & \textbf{Rollout latency (s)} & \textbf{Wait time (s)}  \\
\hline
GSM8K   & 23.45 & 59.50 \\
Search  & 27.98 & 10.99 \\
AMC12   & 70.58 & 15.75 \\
\hline
\end{tabular}
\end{table}

\textbf{Key implication.}
These observations suggest that inefficiency in RLVR systems arises not only from slow rollout and environment phases, but also from strict intra-task sequential execution, which leads to idle time and poor resource utilization. Efficient execution therefore requires treating rollout, environment interaction, and training as \emph{independently schedulable sub-tasks}, and overlapping them across tasks to improve overall system throughput.

\section{Methods}
\begin{figure*}[t]
    \centering
    \includegraphics[width=0.9\textwidth]{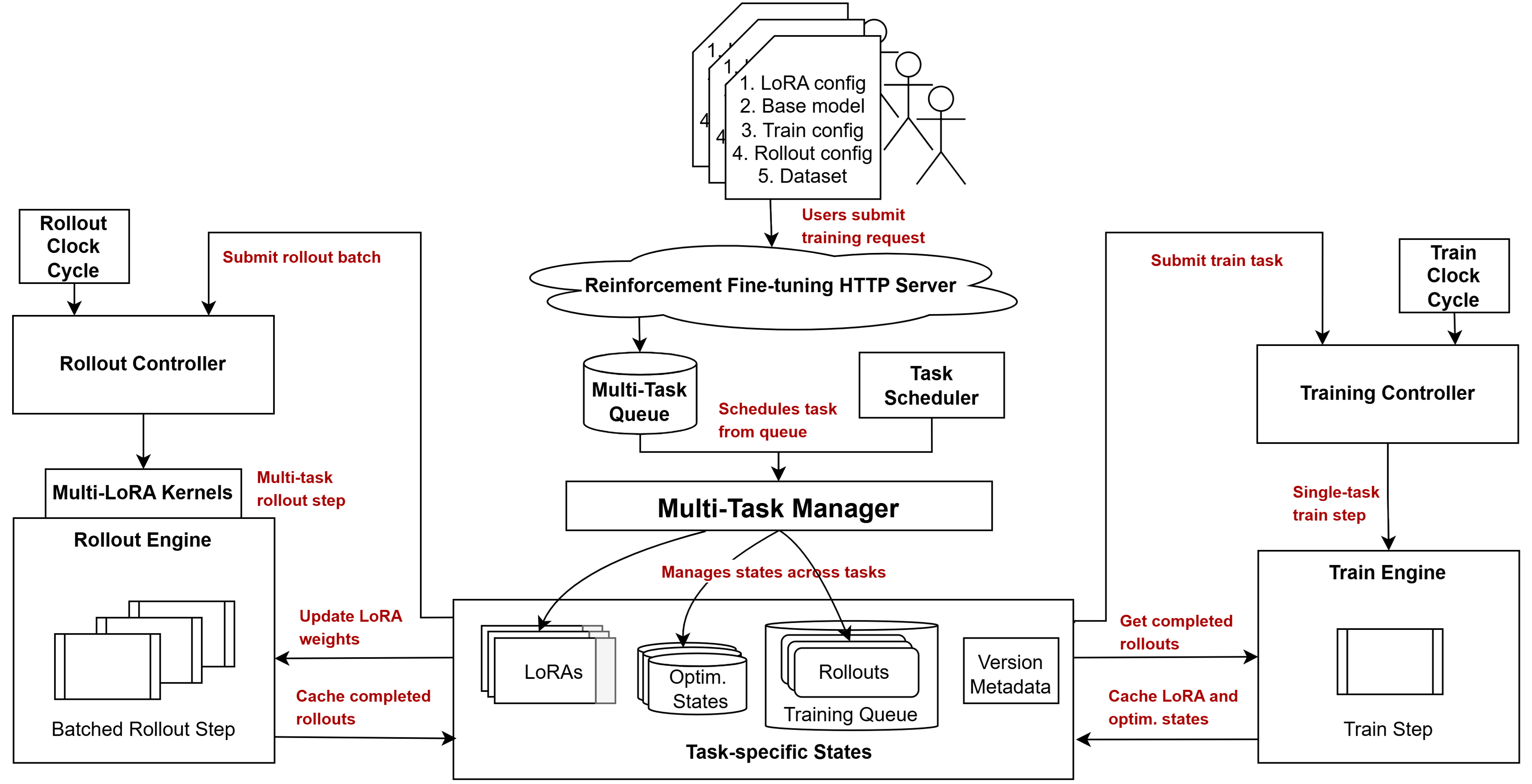}
    \caption{
    \textbf{\textsc{MARLaaS} system architecture.}
    The system consists of a decoupled rollout engine, training engine, and a centralized multi-task manager. Each task maintains independent LoRA parameters and optimizer states. The design enables asynchronous execution while enforcing strict per-task policy versioning.
    }
    \label{fig:design-diagram}
\end{figure*}
We propose \textsc{MARLaaS}, a system for scaling reinforcement learning from verifiable rewards (RLVR) across many concurrent LoRA-adapted tasks. The key design principle is to decouple rollout, environment interactions, and training allowing each task to independently progress when ready.

\subsection{System Overview}

The system is managed by a centralized multi-task manager $\mathcal{M}$ that stores the states required to train each task and coordinates execution between rollout and training engines.

The system, graphically presented in Figure~\ref{fig:design-diagram}, is comprised of of three components: (i) a multi-task manager that maintains versioned state and a global training queue, (ii) a rollout engine that generates trajectories using the next available policy version, and (iii) a training engine that performs policy updates.

\subsection{Multi-Task Manager}

The multi-task manager $\mathcal{M}$ maintains task-specific LoRA parameters, optimizer states, rollout buffers, and policy version metadata for each task $t$. Each task progresses asynchronously with its own version counter.

For each task $t$, $\mathcal{M}$ maintains:
\begin{itemize}
    \setlength\itemsep{0.05em}
    \item LoRA parameters $\theta_t^{(v)}$
    \item Optimizer state $\phi_t^{(v)}$
    \item Version counter $v$
\end{itemize}

The manager additionally maintains a global FIFO buffer $\mathcal{Q}_{buffer}$ storing rollout trajectories paired with their corresponding policy version. 

\subsection{KV-Cache-Aware Scheduling}

We use the rollout engine's memory constraint as an indicator for the number of tasks to be able to schedule concurrently. Each task’s KV-cache footprint is estimated from generation length, batch size, and model configuration. A task is admitted only if the estimated aggregate KV-cache usage remains below a predefined memory budget. Note that this is not a hard constraint as rollouts may queue in the vLLM waiting queue. However, it may cause increases in each task's per-step latency with marginal gains in total throughput. 

\subsection{Asynchronous Execution}

\paragraph{Rollout.}
At each iteration, the rollout engine selects each active task $t$ and retrieves the next unconsumed policy version $(\theta_t^{(v)}, \phi_t^{(v)})$ from $\mathcal{M}$. A trajectory is generated as:
\[
\tau_t^{(v)} \sim \pi_{\theta_t^{(v)}, \phi_t^{(v)}}
\]
The tuple $(t, \tau_t^{(v)}, v)$ is then enqueued into $\mathcal{Q}_{buffer}$. Subsequently, the rollout engine will wait for the next policy version of this task $t$ to be available in $\mathcal{M}$ before processing the next batch of trajectories on the updated policy.

\paragraph{Training.}
The training engine samples $(t, \tau_t^{(v)}, v)$ from the buffer and performs a policy update:
\[
(\theta_t^{(v+1)}, \phi_t^{(v+1)}) \leftarrow \mathrm{PolicyUpdate}(\theta_t^{(v)}, \phi_t^{(v)}, \tau_t^{(v)})
\]
The updated parameters are committed back to $\mathcal{M}$, producing a new policy version. Once committed, the new version becomes eligible for rollout generation and is scheduled by $\mathcal{M}$ for subsequent trajectory generation.

\subsection{Multi-LoRA Rollout with Single-Task Training}

\textsc{MARLaaS} performs rollout generation in parallel across tasks using independent LoRA adapters, while restricting training to a single task update at a time via a centralized queue.

This separation is motivated by the high variance and latency of rollout execution compared to relatively stable and lightweight policy updates. By decoupling these stages, \textsc{MARLaaS} maximizes utilization of rollout compute while avoiding synchronization overhead during training.

\begin{algorithm}[t]
\caption{\textsc{MARLaaS} RLVR Loop}
\begin{algorithmic}[1]

\STATE Initialize multi-task manager $\mathcal{M}$ with $(\theta_t^{(0)}, \phi_t^{(0)}, v)$ for all tasks $t$

\WHILE{system is running}

    \STATE \textbf{Rollout (in parallel)}
    \FOR{each task $t$}
        \IF{$\mathcal{M}.next\_policy(t)$ exists}
            \STATE $(\theta_t^{(v)}, \phi_t^{(v)}) \leftarrow \mathcal{M}.next\_policy(t)$
            \STATE Generate $\tau_t^{(v)} \sim \pi_{\theta_t^{(v)}, \phi_t^{(v)}}$
            \STATE Enqueue $(t, \tau_t^{(v)}, v)$ into $\mathcal{Q}_{buffer}$
        \ENDIF
    \ENDFOR

    \STATE \textbf{Training (in parallel)}
    \IF{$\mathcal{Q}_{buffer}$ not empty}
        \STATE Pop $(t, \tau_t^{(v)}, v)$
        \STATE $(\theta_t^{(v+1)}, \phi_t^{(v+1)}) \leftarrow \mathrm{Policy Update}(\theta_t^{(v)}, \phi_t^{(v)}, \tau_t^{(v)})$
        \STATE Commit updated parameters to $\mathcal{M}$
    \ENDIF

\ENDWHILE

\end{algorithmic}
\end{algorithm}

\section{Experimental Setup}

\paragraph{Environment.}
We evaluate \textsc{MARLaaS} on a two-node Ascend cluster, where each node is comprised of 16 Ascend NPUs (64GB memory each). \textsc{MARLaaS} is implemented on top of AReaL v1.0.1 \cite{fu2025areallargescaleasynchronousreinforcement} and vLLM-Ascend v0.14.0rc1, an Ascend-adapted version of vLLM \cite{kwon2023efficientmemorymanagementlarge}. We leverage the fully sharded data parallel (FSDP) for training \cite{zhao2023pytorchfsdpexperiencesscaling} within AReaL's training engine.

\paragraph{Base Models.}
We use open-weight Qwen models to test the multi-tenant deployment scenarios. Specifically, we evaluate \textsc{Qwen3-0.6B} and \textsc{Qwen3-14B} on a single-node setup, and \textsc{Qwen3-32B} on a two-node setup. This selection allows us to study system behavior across both lightweight and large-scale multi-node regimes.

Since the workloads are rollout-bound, we allocate only the minimum number of accelerators required for training, and dedicate the remaining resources to vLLM for inferencing. Concretely, the 0.6B, 14B, and 32B models require 2, 4, and 16 GPUs for FSDP training, respectively, with all remaining GPUs assigned to vLLM for rollout generation.

\paragraph{Training Algorithm.} We build on the standard GRPO \cite{schulman2017proximalpolicyoptimizationalgorithms} algorithm for our experiments, but our system design is compatible with any policy optimization method that follows the standard rollout-training loop.

\begin{table*}[t]
\centering
\small
\caption{End-to-end training performance across model scales and scheduling regimes on the search agent task.}
\label{tab:perf}
\begin{tabular}{l|cc|cc|cc}
\hline
 & \multicolumn{2}{c|}{0.6B} & \multicolumn{2}{c|}{14B} & \multicolumn{2}{c}{32B} \\
\textbf{Method} 
& Time (hrs) & Steps/hr 
& Time (hrs) & Steps/hr 
& Time (hrs) & Steps/hr \\
\hline
Single-Disaggregated & 18.33 & 54.0 & 24.48 & 39.6 & 25.13 & 38.88 \\
Single-Collocated & 10.64 & 93.6 & 12.70 & 79.2 & 17.98 & 55.62 \\
Multi-LoRA (Sync) & 6.07 & 164.88 & 16.21 & 61.56 & 18.89 & 52.92 \\
\textsc{MARLaas} & \textbf{3.42} & \textbf{292.83} & \textbf{3.72} & \textbf{226.8} & \textbf{9.87} & \textbf{101.30} \\
\hline
\end{tabular}
\end{table*}

\begin{table*}[t]
\centering
\small
\caption{System efficiency and resource utilization under different LoRA scheduling strategies on the search agent task. Metrics report average NPU utilization and idle NPU time (in \%), where higher utilization and lower idle time indicate improved hardware efficiency.}\label{tab:eff}
\begin{tabular}{l|cc|cc|cc}
\hline
 & \multicolumn{2}{c|}{0.6B} & \multicolumn{2}{c|}{14B} & \multicolumn{2}{c}{32B} \\
\textbf{Method} 
& Util (\%) & Idle (\%) 
& Util (\%) & Idle (\%) 
& Util (\%) & Idle (\%) \\
\hline
Single-Disaggregated & 1.56 & 74.18 & 4.45 & 72.52 & 1.58 & 93.18 \\
Single-Collocated & 3.78 & 58.03 & 5.51 & 73.71 & 2.65 & 81.06 \\
Multi-LoRA (Sync) & 1.78 & 85.16 & 3.08 & 86.70 & 1.77 & 87.88 \\
\textsc{MARLaas} & \textbf{6.67} & \textbf{40.52} & \textbf{8.67} & \textbf{40.46} & \textbf{4.35} & \textbf{78.98} \\
\hline
\end{tabular}

\end{table*}

\paragraph{Datasets and Tasks.}
We evaluate \textsc{MARLaaS} on a mixture of reasoning and agentic workloads to capture diverse RLVR training characteristics:

\begin{itemize}
    \item \textbf{GSM8k:} A grade-school math reasoning benchmark requiring multi-step symbolic reasoning \cite{cobbe2021gsm8k}. Use a maximum generation length of 2048 tokens and a batch size of 64.

    \item \textbf{AMC12:} A competition-level math dataset with longer reasoning chains and higher solution complexity \cite{edev2000_amc12_full}. Use a maximum generation length of 4096 tokens and a batch size of 32.

    \item \textbf{Agentic Search:} A tool-augmented reasoning task in which the model interacts with a Wikipedia search API to retrieve and synthesize information. The task is built on top of HotpotQA \cite{yang2018hotpotqadatasetdiverseexplainable}. The workflow may reason and query the Wikipedia search API up to 5 times. We use a separate Qwen3-32B judge model deployed via vLLM on dedicated NPU instances within the same cluster for answer verification. This introduces additional external latency due to tool execution and asynchronous evaluation. We use a maximum generation length of 1024 tokens and a batch size of 32.
\end{itemize}

We deliberately select tasks that are heterogeneous in both reasoning complexity and rollout latency, enabling evaluation of \textsc{MARLaaS} under realistic multi-tenant RLVR workloads.

\paragraph{Baselines.}
We compare \textsc{MARLaaS} against several representative RL training strategies:

\begin{itemize}
    \item \textbf{Single-Disaggregated:} Tasks are trained one at a time with exclusive resource allocation.
    \item \textbf{Single-Collocated:} Rollout and training are executed within a shared resource pool for a single task at a time. This baseline assumes idealized co-location where resource switching is instantaneous and engine re-initialization overhead is negligible. It therefore represents an optimistic upper bound on tightly coupled single-task execution efficiency.
    \item \textbf{Multi-LoRA (Synchronous):} Multiple tasks share rollout and training resources, but synchronization is enforced across tasks.
\end{itemize}

\paragraph{Metrics.}
We evaluate both system efficiency and training performance. Metrics include GPU (AI core) utilization, training throughput (total train steps per hour), average end-to-end job completion time, and resource idle percentage (\%). We measure utilization as average accelerator AI-core utilization percentage reported by profiling tools.

\section{Experiments and Results}

\begin{table*}[t]
\centering
\caption{\textbf{Ablation study of \textsc{MARLaaS}.} We evaluate the contribution of key system components.}
\label{tab:ablation}
\small
\begin{tabular}{lcccc}
\hline
\textbf{Variant} & \textbf{Throughput (steps / hr)} & \textbf{Utilization (\%)} & \textbf{Idle Ratio (\%)} & \textbf{Time (hrs)} \\
\hline
\textsc{MARLaaS} (full) & 255.6 & 22.55 & 17.73 & 1.81 \\
w/o async & 86.4 & 7.04 & 45.01 & 8.13 \\
w/o multi-LoRA & 54.0 & 5.29 & 34.12 & 12.98 \\
\hline
\end{tabular}
\end{table*}

\begin{figure*}[t]
    \centering
    \includegraphics[width=\textwidth]{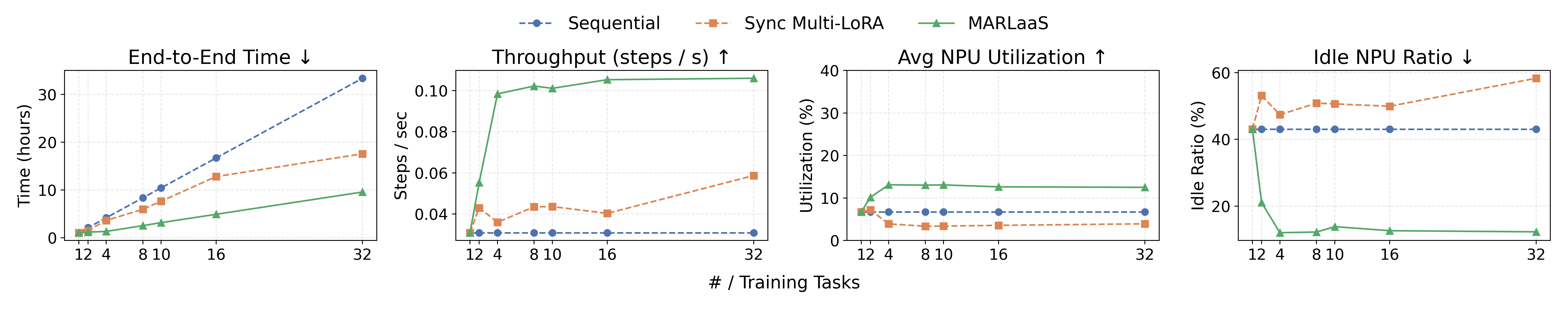}
    
    \caption{
    \textbf{Scaling behavior of \textsc{MARLaaS} under increasing task concurrency.}
    We sweep the number of concurrent RLVR tasks (training \textsc{GSM8K} on \textsc{Qwen3-0.6B}) from 1 to 32. 
    \textsc{MARLaaS} sustains higher utilization and throughput while limiting idle time compared to sequential and synchronous multi-LoRA baselines, demonstrating improved scalability under multi-tenant RLVR workloads.
    }
    
    \label{fig:lora_scaling}
\end{figure*}

\begin{figure*}[t]
    \centering
    \includegraphics[width=0.7\textwidth]{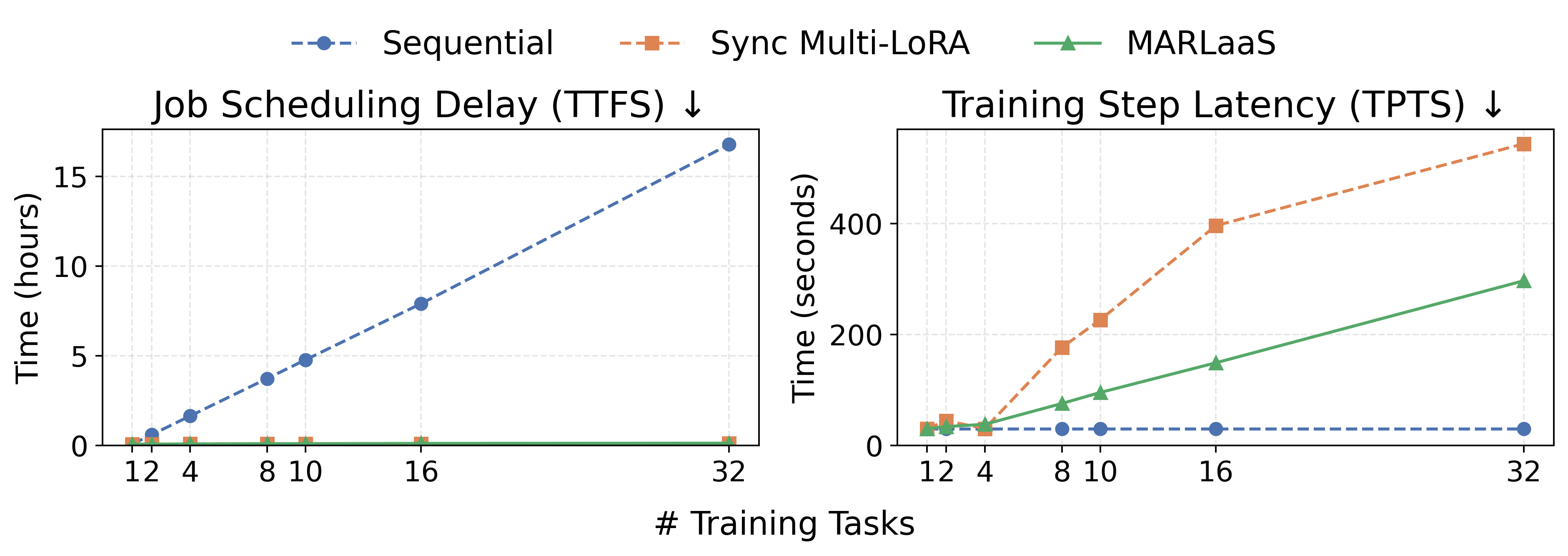}
    \caption{
    \textbf{User-facing latency metrics under increasing concurrency.}
    We compare \textsc{MARLaaS} against sequential and synchronous multi-LoRA baselines as the number of concurrent training tasks increases. 
    We report job scheduling delay (TTFS; time-to-first-step), which captures how quickly a training job begins execution after submission, and training step latency (TPTS; time-per-training-step), which measures per-step iteration latency once training is underway.
    }
    \label{fig:multi-metric-comparison}
\end{figure*}
We evaluate \textsc{MARLaaS} along three axes: (i) end-to-end training performance, (ii) system efficiency and utilization, and (iii) scalability under increasing task concurrency. We further analyze the contribution of key system components through ablation.

Within each experiment, methods share identical model configurations, training hyperparameters, and reward functions. Differences arise solely from scheduling and system design. For multi-tenant workloads, we submit a request to train multiple replicas of each task using the different methods.

\paragraph{End-to-end performance.}
We benchmark \textsc{MARLaaS} against baselines on 100 training steps of ten replicas of the wiki-search agent workload. Table~\ref{tab:perf} reports total wall-clock time and throughput (steps per hour).

\textsc{MARLaaS} consistently outperforms all baselines across model scales, achieving up to $5\times$ higher throughput and substantially reducing training time. This improvement stems from decoupling rollout and training under multi-LoRA concurrency, which eliminates synchronization barriers and increases effective utilization of shared resources.

\paragraph{System efficiency and utilization.}
To understand the source of these gains, Table~\ref{tab:eff} reports average accelerator utilization and idle device time for the same setup.

\textsc{MARLaaS} achieves significantly higher utilization (up to $4.3\times$) while reducing idle time by up to $45\%$ compared to all baselines. This is a direct consequence of the asynchronous execution where multi-LoRA kernels allow cross-task batching of rollouts and the independent nature of interweaving tasks fill up resource bubbles.

Tables~\ref{tab:perf} and \ref{tab:eff} show that performance gains are driven by improved scheduling efficiency and hardware utilization rather than changes to the underlying learning algorithm.

\paragraph{Ablation study.}
To isolate the contribution of key components, we train a \textsc{Qwen3-0.6B} model on ten concurrent AMC12 replicas for one epoch. Results are shown in Table~\ref{tab:ablation}.

Removing asynchronous execution reduces throughput by $\sim66\%$, demonstrating that decoupling rollout and training is the primary driver of efficiency gains. Without asynchrony, the system reintroduces implicit synchronization before each policy update. 

Disabling multi-LoRA rollout further reduces throughput (an additional $\sim13\%$), indicating that concurrency across tasks is also necessary to fully utilize hardware. Without multi-LoRA execution, the system cannot effectively amortize rollout latency across tasks, resulting in lower overall utilization.

These results confirm that both asynchrony and multi-task concurrency are essential and complementary for achieving high system efficiency.

\paragraph{Scalability compared to baselines.}
We evaluate scalability by varying the number of concurrent RLVR tasks from 1 to 32 for 100 training steps each. Figure~\ref{fig:lora_scaling} reports training time, throughput, utilization, and idle ratio.

\textsc{MARLaaS} scales more effectively than sequential and synchronous baselines, maintaining higher utilization as concurrency increases. In particular, throughput continues to improve with additional tasks, while idle time remains comparatively low.

In contrast, synchronous and single-task baselines exhibit diminishing returns. Increased concurrency leads to higher contention and longer waiting times, resulting in underutilized hardware.

Figure~\ref{fig:multi-metric-comparison} reports user-facing latency metrics: job scheduling delay (TTFS) and training step latency (TPTS). TTFS measures the delay between job submission and the first training step, while TPTS captures the latency of each subsequent training iteration.

\textsc{MARLaaS} achieves substantially lower TTFS than the sequential baseline because tasks can begin execution immediately without waiting for prior jobs to complete. Compared to synchronous multi-LoRA training, \textsc{MARLaaS} also maintains lower TPTS by avoiding global synchronization barriers between tasks.

While asynchronous execution introduces modest per-step overhead relative to single-task execution, it significantly improves responsiveness and concurrency under multi-tenant workloads.

These results demonstrate that \textsc{MARLaaS} sustains efficient execution under increasing system load by continuously overlapping rollout and training across tasks.

\section{Discussion}

Our results suggest that multi-tenant RLVR systems are primarily limited by two structural inefficiencies: (i) synchronization overhead between rollout and training, and (ii) redundant model replication across concurrent tasks. \textsc{MARLaaS} addresses these limitations through asynchronous execution and LoRA-based parameter sharing, enabling both temporal and memory-level efficiency improvements.

\paragraph{Asynchronous execution eliminates synchronization bottlenecks.}
Across all settings, asynchronous execution is the primary driver of system efficiency gains. By decoupling rollout generation, environment interaction, and training, \textsc{MARLaaS} removes global synchronization barriers that otherwise force tasks to wait for slower rollouts or coordinated batch completion.

Instead, rollout and training proceed independently and are coordinated only through a shared event queue, allowing compute resources to remain continuously utilized. This eliminates idle periods that arise in tightly coupled pipelines where either rollout or training must block on the other.

As shown in Tables~\ref{tab:eff} and~\ref{tab:ablation}, this design significantly reduces idle device time and increases accelerator utilization. The ablation study further confirms that removing asynchrony leads to the largest degradation in throughput, indicating that synchronization overhead is the dominant system-level inefficiency in prior designs.

We further investigate the tradeoff of user-facing latency metrics in Figure~\ref{fig:multi-metric-comparison}. While asynchronous execution may modestly increase per-step training latency (TPTS) relative to isolated execution, it substantially reduces job scheduling delay (TTFS) by eliminating global synchronization barriers. In multi-tenant RLVR systems, reducing queueing and startup latency is often more important for overall responsiveness than minimizing individual step latency in isolation.

The straggler effect introduced by global synchronization before policy updates becomes even more pronounced when jointly training heterogeneous tasks, as shown in Table~\ref{tab:rollout-tradeoff}. In workloads with highly variable rollout, environment interaction, and training latencies, synchronization overhead can exceed the useful rollout computation itself.

\paragraph{Multi-LoRA sharing improves memory efficiency and task density.}
While asynchrony improves temporal utilization, multi-LoRA parameterization improves spatial efficiency. By sharing a single frozen base model and assigning lightweight LoRA adapters to each task, \textsc{MARLaaS} avoids full model replication across concurrent workloads.

This enables substantially higher task concurrency under fixed hardware and KV-cache constraints. As shown in Table~\ref{tab:perf}, this shared-parameter design reduces end-to-end runtime by approximately $1.8\times$--$6.5\times$ when training 10 concurrent tasks, compared to baselines without multi-LoRA sharing.

Importantly, these gains arise from improved hardware utilization rather than changes to the underlying optimization algorithm, and they complement the benefits of asynchronous execution.

\paragraph{Impact of increasing concurrency.}
As shown in Figure~\ref{fig:lora_scaling}, efficiency gains are greatest when scaling from single-task execution to moderate concurrency (up to 4 tasks in this experiment). In this regime, additional tasks effectively fill idle rollout and training capacity, leading to substantial improvements in throughput and utilization.

Beyond this point, performance gains diminish. The dominant bottleneck shifts from idle compute and synchronization overhead at low concurrency to contention for shared resources, including KV-cache capacity and serialized policy updates.

Although throughput continues to improve with higher concurrency, the benefit decreases while user-facing latency metrics such as TTFS and TPTS begin to increase (Figure~\ref{fig:multi-metric-comparison}). This highlights a broader scheduling tradeoff in multi-tenant RLVR systems between maximizing overall throughput and maintaining low per-task latency.

An important future direction is adaptive concurrency control that dynamically balances throughput, TTFS, and TPTS under changing workload conditions, resource availability, and user requirements.
\section{Conclusion}

We introduced MARLaaS, a multi-tenant asynchronous RLVR training system that decouples rollout generation, environment interaction, and policy optimization while enabling shared LoRA-based parameterization across tasks. Our results show that eliminating synchronization bottlenecks and improving resource overlap leads to substantial gains in hardware utilization and end-to-end training time, without sacrificing learning performance. These findings highlight the importance of asynchronous execution and memory-aware scheduling in scaling RLVR systems, and suggest promising directions for future multi-tenant reinforcement learning infrastructure.

\section*{Limitations}
While \textsc{MARLaaS} improves utilization and concurrency for RLVR workloads, several limitations remain. First, the current design serializes policy updates through a single-task training engine, which eventually becomes a bottleneck under high concurrency. Second, KV-cache capacity still fundamentally limits rollout scalability for long-context or heavily agentic tasks. Future work should explore distributed training execution, adaptive scheduling policies, and dynamic resource allocation strategies to adapt to evolving training-step characteristics and user objectives. 

At present, the system does not include adaptive mechanisms for task prioritization or dynamic resource allocation, both of which could further improve efficiency in real-world deployments. Additionally, \textsc{MARLaaS} has been evaluated on a range of heterogeneous datasets across a range of model sizes. While we report consistent findings, we acknowledge that the results remain sensitive to workload and experimental characteristics, and the generalizability of our approach warrants further validation.

\bibliography{custom}

\end{document}